\documentclass[twocolumn]{aastex62}

\usepackage{subfigure,epsfig,amsfonts}
\usepackage{natbib}
\usepackage{amsmath}
\usepackage{amssymb}
\usepackage{amsthm}
\usepackage{graphicx}
\usepackage{commath}
\usepackage{url}

\DeclareMathOperator{\sgn}{sgn}
\newcommand{\Kepler}{\emph{Kepler\ }}
\newcommand{\Gaia}{\emph{Gaia\ }}

\usepackage{comment}

\shorttitle{CKS VIII}

\begin{document}

\title{CKS VIII: Eccentricities of Kepler Planets and \\Tentative Evidence of a High Metallicity Preference for Small Eccentric Planets}
\correspondingauthor{Sean M. Mills}
\email{sean.martin.mills@gmail.com}

\author{Sean M. Mills}
\affiliation{California Institute of Technology, Department of Astronomy\\
1200 East California Blvd, Pasadena, CA 91125, USA \\
}

\author{Andrew W. Howard}
\affiliation{California Institute of Technology, Department of Astronomy\\
1200 East California Blvd, Pasadena, CA 91125, USA \\
}

\author{Erik A. Petigura}
\affiliation{California Institute of Technology, Department of Astronomy\\
1200 East California Blvd, Pasadena, CA 91125, USA \\
}

\author{Benjamin J. Fulton}
\affiliation{California Institute of Technology, Department of Astronomy\\
1200 East California Blvd, Pasadena, CA 91125, USA \\
}
\affiliation{IPAC-NASA Exoplanet Science Institute \\
Pasadena, CA 91125, USA \\
}

\author{Howard Isaacson}
\affiliation{Department of Astronomy, University of California\\
510 Campbell Hall, Berkeley, CA 94720, USA\\
}

\author{Lauren M. Weiss}
\affiliation{Institute for Astronomy, University of Hawaii \\
2680 Woodlawn Drive, Honolulu, HI 96822, USA\\
}
\affiliation{Parrent Fellow}

\shorttitle{CKS VIII: Eccentric Planets Prefer Metal-Rich Hosts}
\shortauthors{Mills et al.}

\begin{abstract}

Characterizing the dependence of the orbital architectures and formation environments on the eccentricity distribution of planets is vital for understanding planet formation. In this work, we perform statistical eccentricity studies of transiting exoplanets using transit durations measured via \emph{Kepler} combined with precise and accurate stellar radii from the California-\emph{Kepler} Survey and \emph{Gaia}. Compared to previous works that characterized the eccentricity distribution from transit durations, our analysis benefits from both high precision stellar radii ($\sim$3\%) and a large sample of $\sim$1000 planets. We observe that that systems with only a single observed transiting planet have a higher mean eccentricity ($\bar{e} \sim 0.21$) than systems with multiple transiting planets ($\bar{e} \sim 0.05$), in agreement with previous studies. We confirm the preference for high and low eccentricity subpopulations among the singly transiting systems. Finally, we show suggestive new evidence that high $e$ planets in the \emph{Kepler} sample are preferentially found around high metallicity ([Fe/H] $>0$) stars. We conclude by discussing the implications on planetary formation theories.

\end{abstract}

\section{Introduction}
\label{sec:intro}

The exquisite photometric data from the \Kepler mission has revolutionized our knowledge of exoplanet demographics. Studies revealed the occurrence rate of planets as a function of orbital period and planet size \citep{2013PNAS..11019273P,2018ApJ...860..101Z}, the low mutual inclination of multiplanet \Kepler systems \citep{2014ApJ...790..146F,2018ApJ...860..101Z}, and the orbital period ratio distribution of the \Kepler planets \citep{2014ApJ...790..146F,2015MNRAS.448.1956S}. Recently, large spectroscopic surveys enabled improved measurements of \Kepler planet host star properties \citep[e.g.,][]{2015ApJS..220...19D,2017AJ....154..108J}. These improved stellar properties revealed new details in the planet population, such as the bimodal distribution of planets between 1 and 4 $R_\oplus$ \citep{2017AJ....154..109F}, and their eccentricity ($e$) distribution \citep{2016PNAS..11311431X}. Measuring the eccentricity distributions of different planet populations is important because they are relics of the processes which occurred during the epoch of planet formation and migration.

The eccentricity of planets detected with radial velocity can be measured via the shape of the Keplerian signal, and a large range of eccentricities have been observed \citep{2015ARA&A..53..409W}. However, due to the difficulty in constraining planet eccentricity at low radial velocity (RV) amplitudes \citep{2008ApJ...685..553S}, generally only in giant planet systems with high signal-to-noise (S/N) RV data can eccentricity be well constrained. 

On the other hand, exoplanet eccentricity can be determined in systems where multiple planets transit via a dynamical analysis. The gravitational interaction among planets causes transit timing variations (TTVs), which are sensitive to the orbital period ratios, masses, and eccentricities of the planets \citep{2005MNRAS.359..567A,2012ApJ...761..122L}. 
However, measurements of eccentricity by TTVs often suffer from two limitations: (1) it nearly always requires multi-transiting systems and (2) the eccentricity is often degenerate with planet mass \citep{2012ApJ...761..122L}. Thus the \Kepler data can uniquely determine eccentricities only in rare cases where this degeneracy is broken \citep{2015ApJ...802..116D}. However, a planet's eccentricity also subtly affects the duration of a planet's transit regardless of any dynamical interactions. 

The duration of a planetary transit is determined by the length of the transit chord across the face of its host star divided by the planet's orbital velocity. The transit chord length is given by the radius of the star ($R_\star$) and the impact parameter ($b$) of the transit. The velocity of a planet on a circular orbit is uniquely determined by the planet's orbital period and the stellar mass, assuming $M_p \ll M_\star$. However, an eccentric planet's velocity depends additionally on the eccentricity and the phase of the planet during transit (since the orbital velocity is not constant throughout the orbit). This results in a degeneracy between $b$ and $e$. Because the impact parameter also affects the transit shape \citep{2010exop.book...55W}, careful transit modeling may uniquely determine $b$ and therefore also $e$. Caution must be taken to account for transit timing variations (TTVs), which may also alter the apparent shape of transits phased at constant period \citep{2014MNRAS.440.2164K}. This has been done for a subset of \Kepler planets which have precisely characterized stellar hosts from asteroseismology \citep{2015ApJ...808..126V,2019AJ....157...61V}, resulting in a measurement of the low ($\lesssim 0.06$) eccentricities of $\lesssim 4 R_\oplus$ planet pairs and confirming previous results of high mean eccentricity ($e \sim 0.2$) for single transiting planet systems \citep{2016PNAS..11311431X}. An alternative approach to the precise characterization of individual planet transits is to use a statistical methodology which eliminates the need for individually measured impact parameters \citep{2008ApJ...678.1407F}. This technique has previously been applied to the \Kepler data with varying levels of stellar host property precision \citep{2011ApJS..197....1M,2012MNRAS.425..757K,2014PASP..126...34P,2016PNAS..11311431X}. 
The most recent result, \citet[][X16 hereafter]{2016PNAS..11311431X}, reveals that single transiting planet systems from \Kepler are drawn from a significantly broader distribution of eccentricities than the multiple transiting planet systems. A similar effect has been observed in systems of giant planets in single and multiple planet systems detected via RVs \citep{2013Sci...340..572H,2015PNAS..112...20L}.

The work in this paper combines the unprecedented population-wide precision and accuracy of the CKS-\Gaia stellar sample \citep{2018AJ....156..264F} with the most recent \Kepler data release \citep[DR25;][]{2016AJ....152..158T} to improve our knowledge of the eccentricities of the \Kepler planets. We simulate populations of exoplanet systems with different eccentricity distributions and compare the resulting transit duration distribution to the distribution of durations observed with \Kepler to determine the most likely eccentricity of this population. We use these results to search for trends in planet and host star properties as a function of eccentricity.

\section{Methods}
\label{sec:methods}

The population approach to measure eccentricity distributions from transit durations assumes that the three-dimensional orientation of exoplanet systems is isotropic. Therefore the distribution of impact parameters will be uniform, subject to the observational bias that planets with high impact parameters have shorter and shallower transits and are thus less likely to be detected. Thus by assuming a given population of planets has randomly oriented invariant planes, the deviation from the expected distribution of observed durations reveals the eccentricity distribution of the population. A highly eccentric population will have more short duration transits than a circular population due to the higher orbital velocities near pericenter combined with the increased transit probability due to the decreased planet-to-star separation \citep{2008ApJ...679.1566B}. Numerically integrating Eq. 16 from \citet{2008ApJ...679.1566B} reveals that only $\sim$10-20\% of transit durations will be longer than expected for any eccentricity $\sim$0.1-0.8 randomly observed from different orientations many times. This long-duration subset also favors durations just longer than the $b=0$, $e=0$ geometry, and can therefore be difficult to distinguish from a circular orbit when measurement errors are considered. If, however, a planet's transit is longer than expected for the $b=0$, $e=0$ case even when accounting for all uncertainties, a high eccentricity with transit near apocenter is a unique conclusion. On the other hand, the plentiful short-duration transits may each individually be explained by a high $b$ rather than $e$. But by enforcing the assumption of random viewing orientations for a planet population, the population's high $e$ distribution may be determined by the surplus of these short duration events over that expected from an $e=0$ population. This method is also not biased by transit timing variations since time-shifts in the transits do not change the average transit duration.

To construct and compare our simulated populations to the observed \Kepler population, we closely follow the methodology described in X16. We summarize the steps as follows\footnote{Source code for the analysis is available for download at \url{https://github.com/smmills/CKS_Eccentricities}.}:

\subsection{Single Transiting Planet Systems}
\label{sec:methodssingles}

We adopt a truncated Rayleigh distribution of eccentricities in our model since it is commonly used for eccentricities \citep{2008ApJ...678.1407F,2011ApJS..197....1M,2014ApJ...790..146F,2016PNAS..11311431X} and thus readily comparable to literature results. This choice is physically motivated \citep{2016ApJ...820...93S}, and has support over the allowed range [0,1). The probability distribution function (PDF) of the truncated Rayleigh (TR) is:

\begin{equation}
\mathrm{PDF}_{\mathrm{TR}}(\sigma_e) = \left\{
	\begin{array}{ll}
		\frac{e}{\sigma_e^2} \exp(- \frac{e^2}{2\sigma_e^2} ) & 0 <= e < 1 \\
		0 & \mathrm{otherwise}
	\end{array}
	\right .
\end{equation}

For each system, we (1) draw an eccentricity from a the truncated Rayleigh distribution with a specified width parameter, $\sigma_e$. (2) We draw an invariant plane and pericenter ($\omega$) direction of the system at random. We exclude cases where the planet is guaranteed not to transit. (3) We compute the resulting transit duration. If the duration is unphysical due to, e.g., the pericenter of the planet hitting the star, we begin again from step 1. We also consider if the resulting transit would be detectable by computing the expected S/N of the simulated transit as $\sqrt{d_{sim}/d_{obs}} * SN_{obs}$, where $d$ is the transit duration with subscripts for the simulated ($sim$) and observed ($obs$) cases, and $SN_{obs}$ is the measured S/N of the Kepler Object of Interest (KOI). If the simulated transit S/N does not meet the threshold for inclusion in the \Kepler catalog ($>$7.1), we start again from step 1. (4) We then compute the duration ratio ($r$) of the simulated transit to a transit with $e=0$ and $b=0$, as well as the observed duration ratio to a nominal $e=0$, $b=0$ transit. The likelihood of the observed transit duration ratio is computed as 
\begin{equation}
\mathcal{L}_r = \exp \bigg(- \big(\frac{r_{sim}-r_{obs}}{\sigma_{obs}}\big)^2 \bigg),
\end{equation}
where $\sigma_{obs}$ is the uncertainty of $r_{obs}$. 

These steps are repeated 10,000 times for each KOI to probe the distribution of simulated durations for a given system and eccentricity distribution. The likelihoods are then multiplied to give a total likelihood for the $e$ distribution given the observed transit duration. This procedure generates a Monte Carlo approximation of the integral 
\begin{equation}
\mathcal{L}_{\sigma_e} = \int_r P(r_{sim} | \sigma_e) P(r_{obs}) \mathrm{d} r.
\end{equation}
We repeat this procedure 100 times and average the values of the trials at each $\sigma_e$ for a large grid of $\sigma_e$s to determine the overall $\mathcal{L}(\sigma_e)$ for a given KOI.

Our catalog of KOIs includes all planets and planet candidates from \Kepler Data Release 25 \citep{2016AJ....152..158T}. To prevent spurious results from binary stars and other false positives, we remove any KOIs from our sample with $R_p > 15$ or that has either a False Positive Flag of 1 or a False Positive Probability (FPP) $>0.5$ in the \citet{2016ApJ...822...86M} catalog. We find the results are not sensitive to the exact FPP cutoff chosen. We also consider the effect of changing the S/N cutoff of 7.1 to reduce the number of false positives caused by noise. However, we find our results do not change significantly when higher S/N cutoffs are chosen (10 or 12), and therefore retain 7.1 as the nominal cutoff.  

We restrict our sample to $ 0.5 R_\odot < R_\star < 2.0 R_\odot$, which includes the majority of \Kepler targets and reduces the chance of calibration errors. We remove any systems whose uncertainty in $R_p$ exceeds $0.02$ $R_\star$ -- this cutoff removes the handful of systems which may be unreliably measured in the \Kepler dataset due to limb-darkening degeneracies. Finally, we remove stars that \citet{2018AJ....156..264F} identify as having $\ge 5\%$ contamination from nearby sources, as these stars may sometimes have biased or incorrect radii. The exact value of these cutoffs does not significantly affect our results. We are left with 439 singly transiting KOIs whose $\mathcal{L}_{\sigma_e}$ we sum to get the overall population likelihood as a function of $\sigma_e$. 

Due to the possibility of unaccounted for systematics, we conservatively report 2-$\sigma$ equivalent uncertainties on $e$ values.

\subsection{Multiple Transiting Planet Systems}
\label{sec:methodsmultis}

Our approach is fundamentally the same for the systems with two or more observed transiting planets as for single transiting planet KOIs, with the exception that we must also take into account the mutual inclination dispersion among the multiplanet systems which will also affect the transit durations. For instance, for extremely coplanar planetary systems by geometric arguments alone an interior planet must have a smaller impact parameter than an exterior planet. This is not true for a significantly mutually inclined system. Thus the mutual inclination of the systems affects the distribution of assumed underlying impact parameters, which is no longer uniform. To address this we modify step (2) above by drawing the mutual inclination and thus orbital plane of each planet in a system from a Gaussian distribution with width $\sigma_i$ centered around the invariant plane. When we reach step (3), we check that every planet in the system meets the S/N threshold and restart the entire system from step (1) if any do not.

We then perform a grid-search over both $\sigma_e$ and $\sigma_i$ to determine the likelihood surface in terms of both mutual inclination and eccentricity dispersion. To make the process more computationally tenable, for theses systems we use 20 trials of the 10,000 point Monte Carlo integrals computing
\begin{equation}
\mathcal{L}_{\sigma_e,\sigma_i} = \int_r P(r_{sim} | \sigma_e, \sigma_i) P(r_{obs}) \mathrm{d} r
\end{equation}
for each system. The same cutoffs are then applied as for the singles (see \S\ref{sec:methodssingles}) before computing the full population likelihood as a function of $\sigma_e$ and $\sigma_i$. The multiplanet sample consists of 870 KOIs.

\section{Single Planet Results}

\begin{figure}
\centerline{
\includegraphics[scale=0.6]{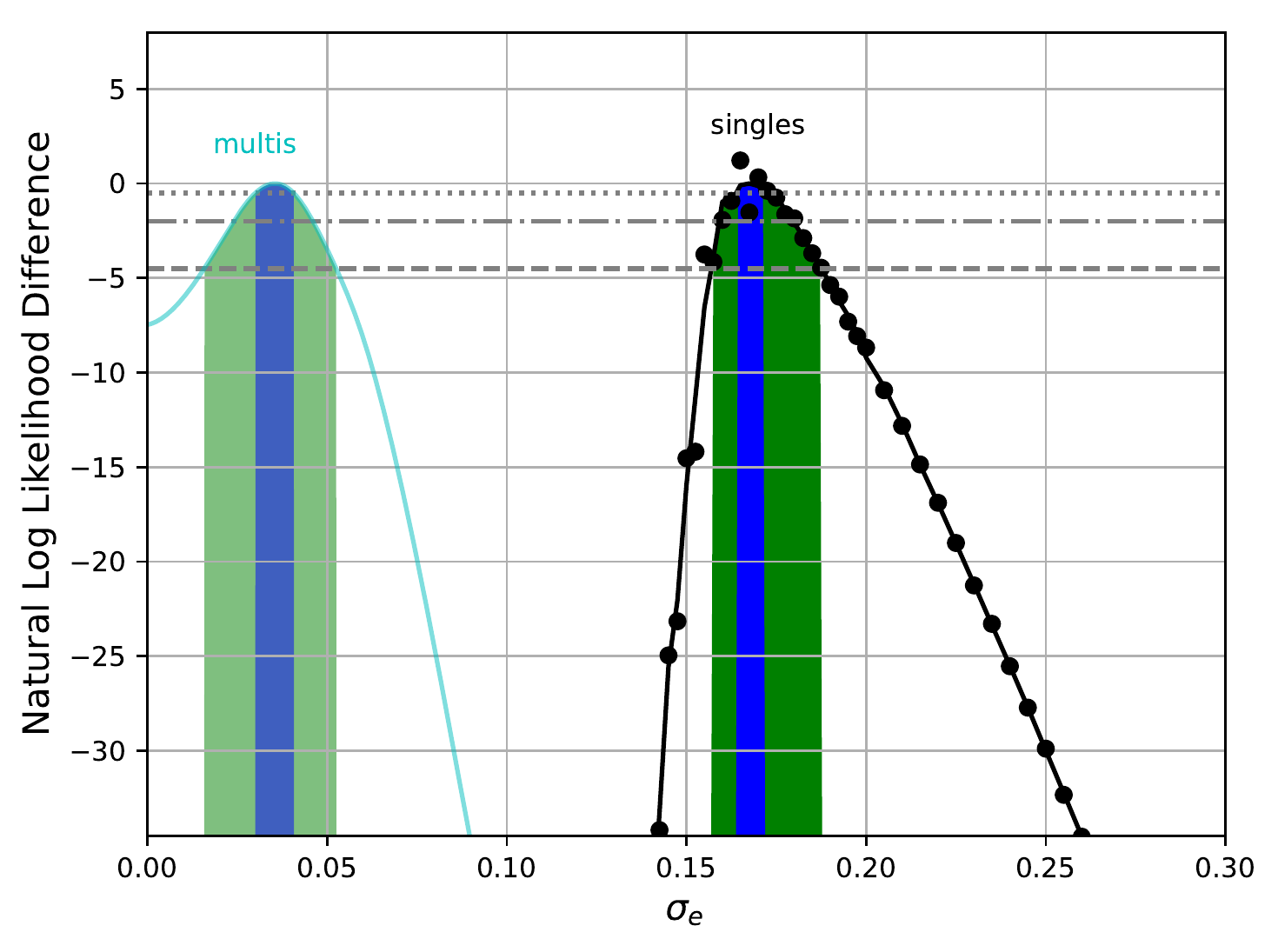}
}
\caption{ The natural log-likelihood of the vetted population of singly transiting planet systems as a function of $\sigma_e$. The data and the Savitzky-Golay interpolation are shown in black. Dotted, dot-dashed, and dashed horizontal lines indicate nominal 1-, 2-, and 3-$\sigma$ likelihood differences respectively. The 1- and 3-$\sigma$ confidence intervals are shaded opaque blue and green respectively. For comparison, the distribution of $\sigma_e$ likelihoods of the multiplanet systems at the best-fit mutual inclination is shown in cyan, with 1- and 3-$\sigma$ ranges shaded in translucent blue and green respectively. This corresponds to a horizontal slice in Fig.~\ref{fig:multis2d} at $\sigma_i = 0.043$. 
}
\label{fig:allfit}
\end{figure}

We find that the single-planet systems are best fit with $\sigma_e=0.167^{+0.013}_{-0.008}$ at the 95\% confidence level (see Fig.~\ref{fig:allfit}). The maximum and uncertainties are found by interpolating likelihoods from the grid of $\sigma_e$ with a degree 3 Savitzky-Golay filter \citep{1964AnaCh..36.1627S} using the 7 nearest neighbors to each point. The best-fit value is driven by the balance between the majority of the population of planets whose transit durations are consistent with low eccentricities, and roughly a dozen systems which disfavor $e=0$ to moderate to high significance ($2-40\sigma$). We list systems which individually strongly suggest a high eccentricity in Table~\ref{table:highesingles}. These systems have longer transit durations than possible with a circular orbit (see \S~\ref{sec:methods}) and are unlikely to have arisen by chance given the uncertainties. We also hand-inspected each of these lightcurves and found nothing anomalous about these candidates.

\begin{deluxetable}{rclll}
\tablecaption{High Eccentricity Planets\label{table:highesingles}}
\tablehead{  \colhead{KOI} & \colhead{High $\sigma_e$ } & \colhead{$d_{obs}$ (d)} & \colhead{$d_{circ}$ (d)} & \colhead{$R_\oplus$} \\ 
		& \colhead{Preference} & &  &\\
}
\startdata
2046.01 & 38$\sigma$ & 0.66 & 0.35 & 2.7  \\
144.01  & 21$\sigma$ & 0.15 & 0.11 & 3.1  \\
2698.01 & 19$\sigma$ & 0.61 & 0.39 & 3.4   \\
2904.01  & 6.9$\sigma$ & 0.41& 0.30 & 2.2   \\
3678.01  & 5.3$\sigma$ & 0.45 & 0.41 & 7.9   \\
333.01  & 5.3$\sigma$ & 0.26 & 0.22 & 2.7   \\
4156.01$^a$  & 3.6$\sigma$ & 0.24 & 0.20 & 1.8   \\
\hline
\enddata
\tablenotetext{}{Single transiting planet systems which favor high $e$ at $>3.5\sigma$ and multiply transiting systems which favor high $e$ at $>3.7\sigma$. These cutoffs were chosen so that the expected number of false positives for each sample is less than 1.
}
\tablenotetext{a}{The measured duration of KOI 4156.01 is 6-$\sigma$ shorter in Data Release 24 compared to Data Release 25, so we view this candidate with caution.}
\end{deluxetable}

\begin{figure}
\centerline{
\includegraphics[scale=0.6]{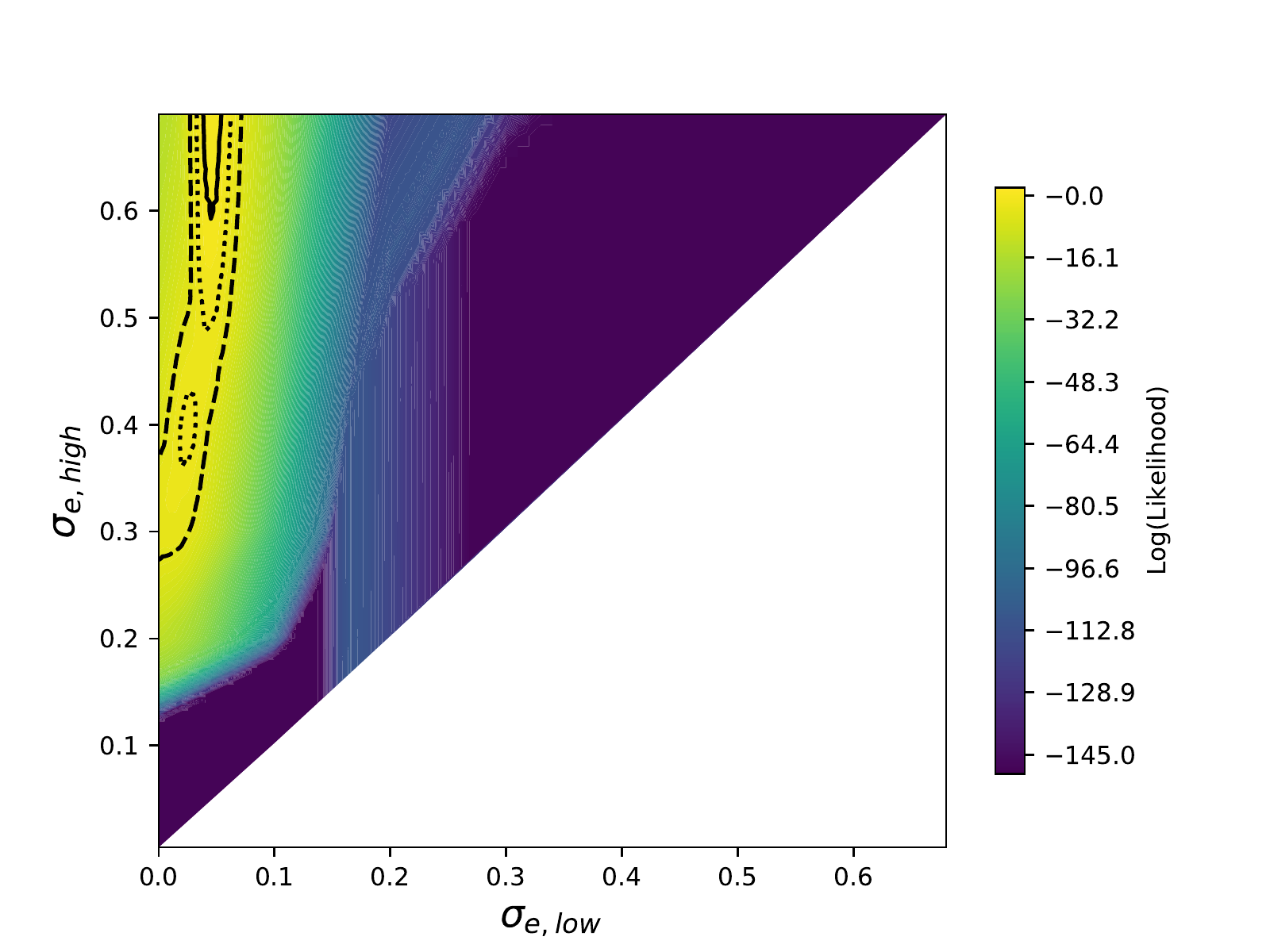}
}
\caption{ The likelihood surface of the two population model as a function of the $\sigma_{e,low}$ and $\sigma_{e,high}$ for the singly-transiting planet systems. Solid, dotted, and dashed contours indicate 1-, 2-, and 3-$\sigma$ contours respectively. The greatest likelihood value has 69\% of the systems in the low eccentricity distribution. 
}
\label{fig:singles2D}
\end{figure}

The existence of only a few systems which strongly suggest high eccentricity motivated us to consider a two-population eccentricity model, as also investigated in X16 and \cite{2019AJ....157...61V}. Our model is a simple combination of two Rayleigh distributions $\sigma_{e,low}$ and $\sigma_{e,high}$, and the fraction of systems in the low $e$ distribution ($f$). To fit for the $\sigma_e$ values for each population, we chose a a pair of $\sigma_e$ values from our grid search, sort the planets by their best-fit $\sigma_e$ values, and find the division of the sorted list such that the likelihood is maximized. We then iterate over all possible  $\sigma_{e,low}$ and $\sigma_{e,high}$ pairs, and compare the highest likelihood values for all pairs. This results in a triangular 2D likelihood surface with a ridge peak indicating two populations with $\sigma_{e,low} \lesssim 0.05$ and $\sigma_{e,high} \gtrsim 0.3$ best fit the data (Fig.~\ref{fig:singles2D}). In the best-fit solution, 69\% of systems preferred the low-eccentricity distribution. We note that at high $\sigma_e$ the eccentricity distribution changes very little due to the cutoff at $e=1$. This causes the weak dependency on $\sigma_{e,high}$ for $\sigma_{e,high} \gtrsim 0.3$ seen in Fig.~\ref{fig:singles2D} and explains our decision to truncate our search at $\sigma_e=0.7$ ($\sigma_e$ can be arbitrarily large since the distribution is truncated at $e=1$, but the distributions become unphysical at high $\sigma_e$).

We compare the likelihood of the single-population and two-population model to determine if the two-population model is warranted. The introduction of the two additional free parameters (fraction of systems in the low-e population, and $\sigma_e$ for the low-e population) increased the natural log likelihood by 108, a $\Delta$ AIC \citep[Akaike Information Criteria;][]{1974ITAC...19..716A} of 212, strongly favoring a 2-population model \citep{burnham2003model}. We note that this is not trivial due to the small number of systems with a confidently detected high eccentricity, because for a single high eccentricity population the expected number of systems with anomalously long eccentricities compared to a circular distribution would be small due to the decreased transit probability \citep{2008ApJ...679.1566B}. Our results are similar to the two-population results in X16, and for a more thorough analysis of the two-population models and eccentricity distributions see \cite{2019AJ....157...61V}.

\begin{figure*}
\centerline{
\includegraphics[scale=0.4]{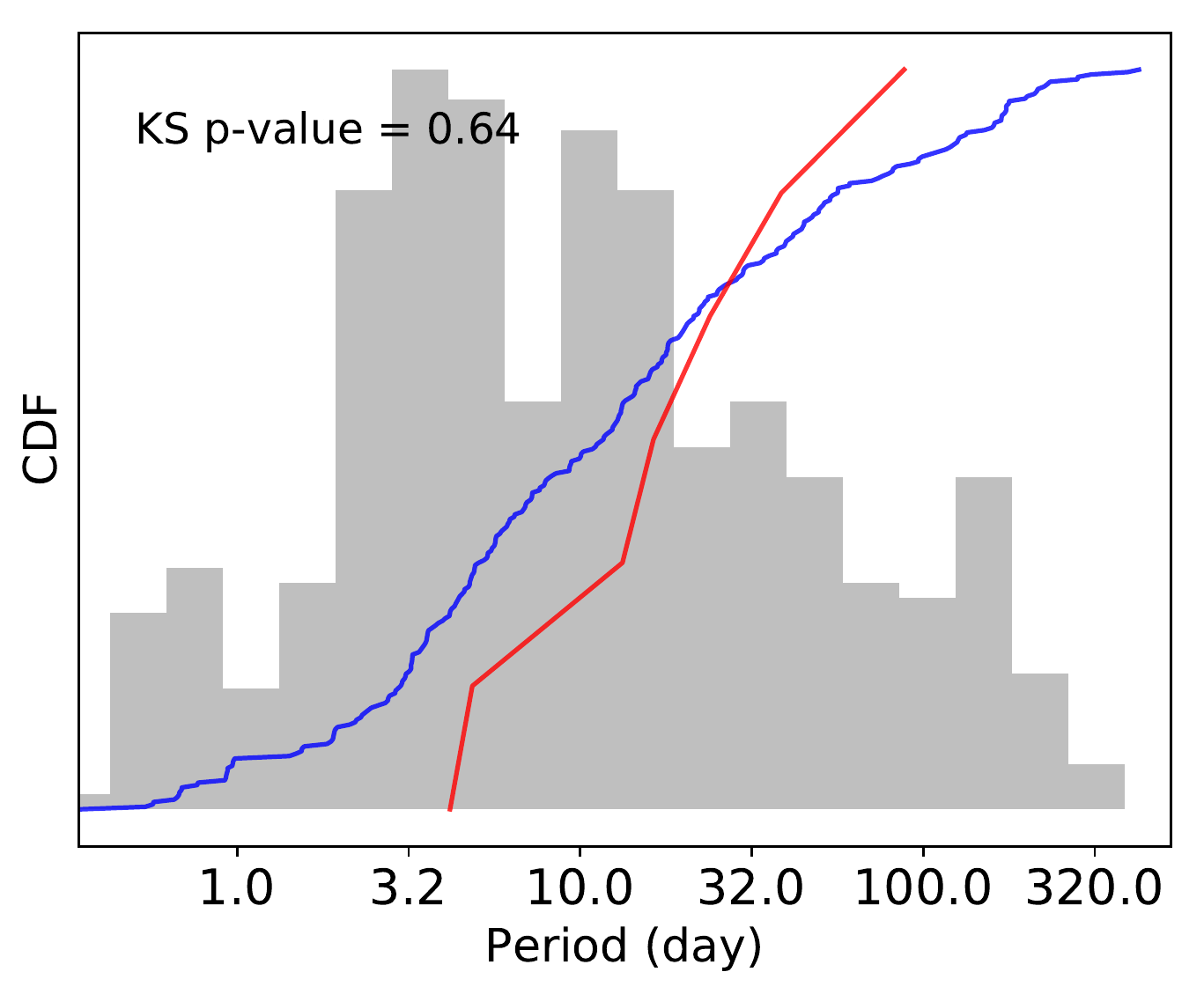}
\includegraphics[scale=0.4]{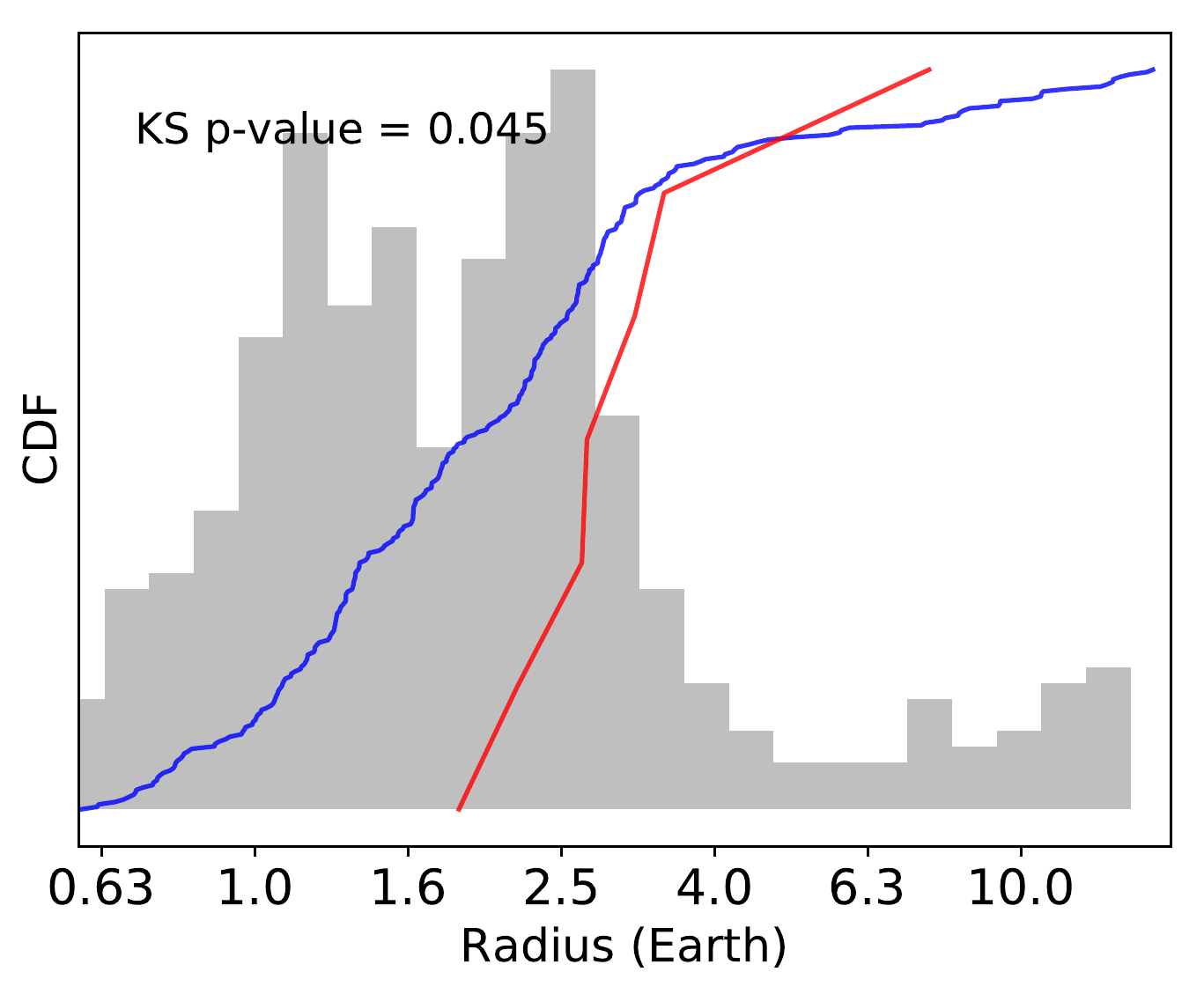}
}
\centerline{
\includegraphics[scale=0.4]{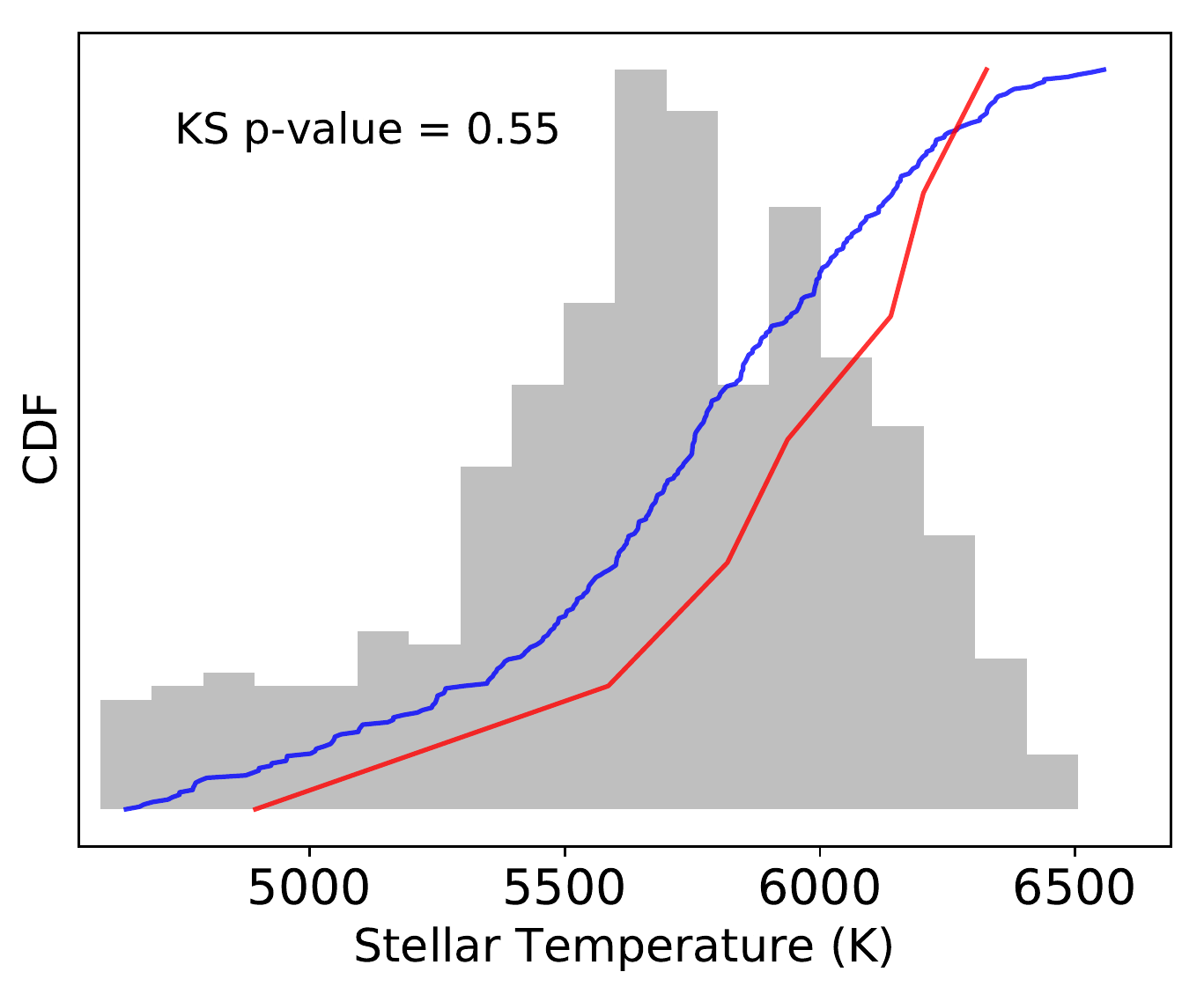}
\includegraphics[scale=0.4]{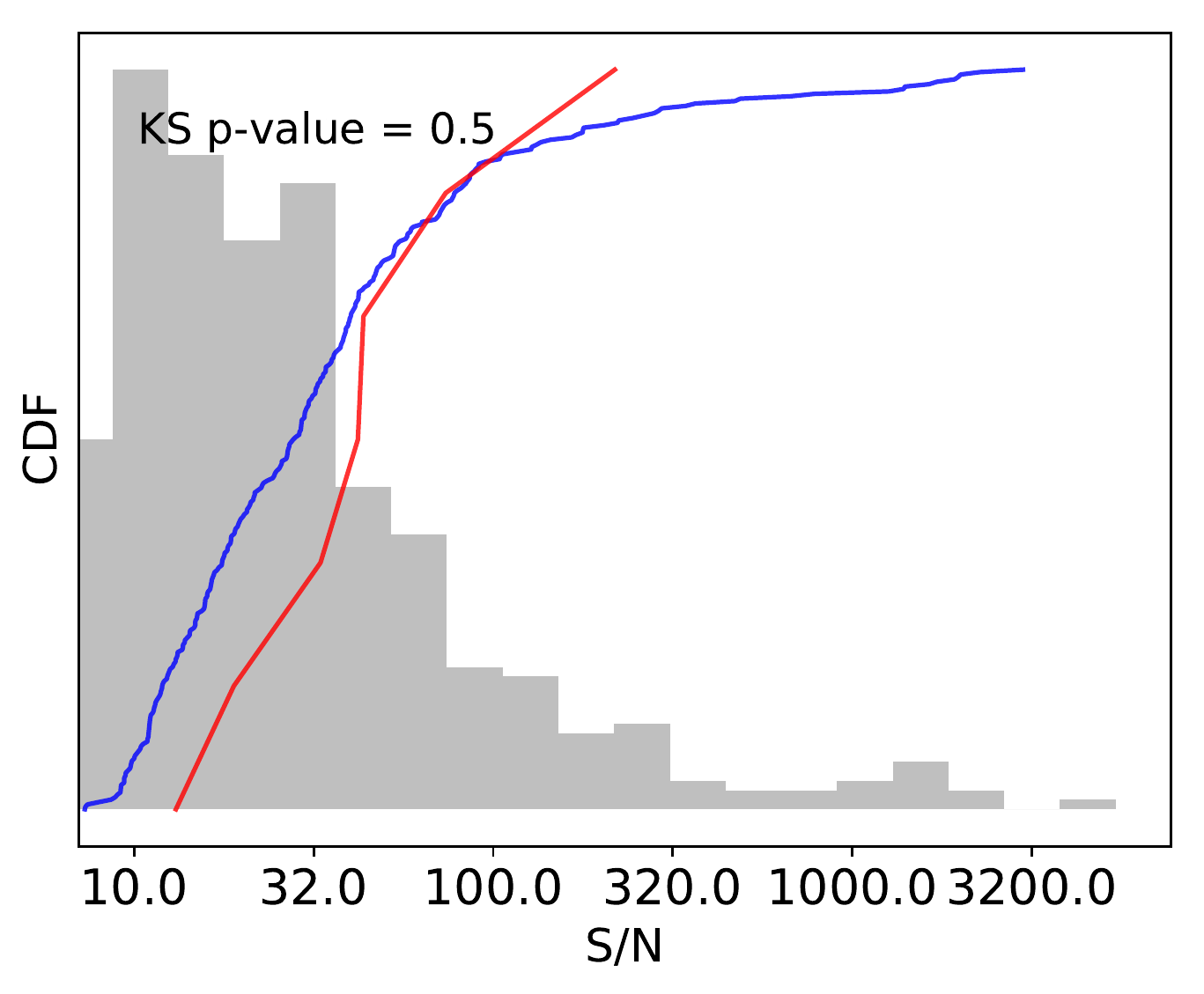}
}
\caption{ The CDFs of the population of singly-transiting planet systems which support low eccentricities (blue) and the 5\% of systems which most strongly favor high eccentricities (red) for various planetary and stellar properties. The PDF of the entire sample is shown in gray. 
}
\label{fig:singlesall}
\end{figure*}

\begin{figure*}
\centerline{
\includegraphics[scale=0.4]{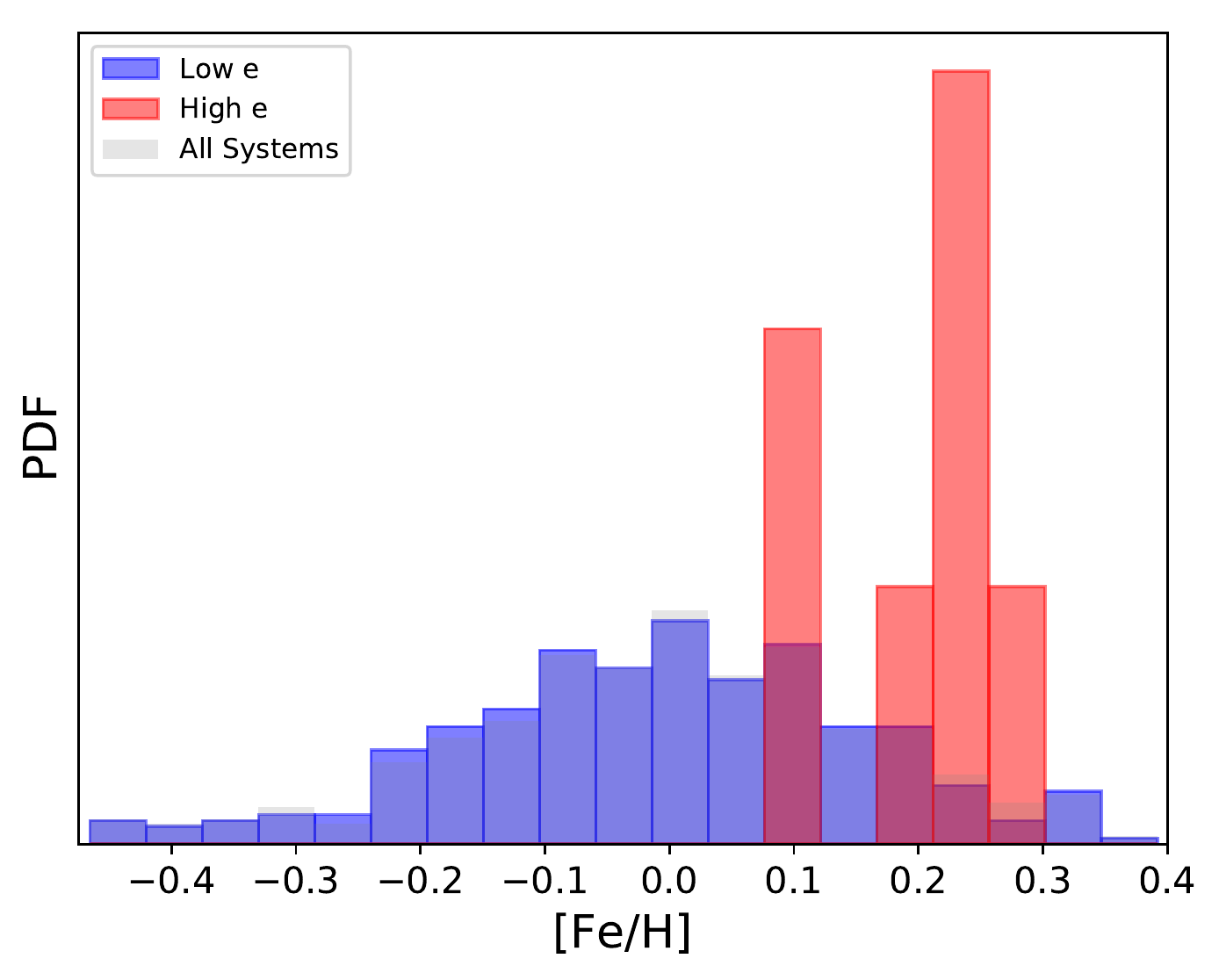} 
\includegraphics[scale=0.4]{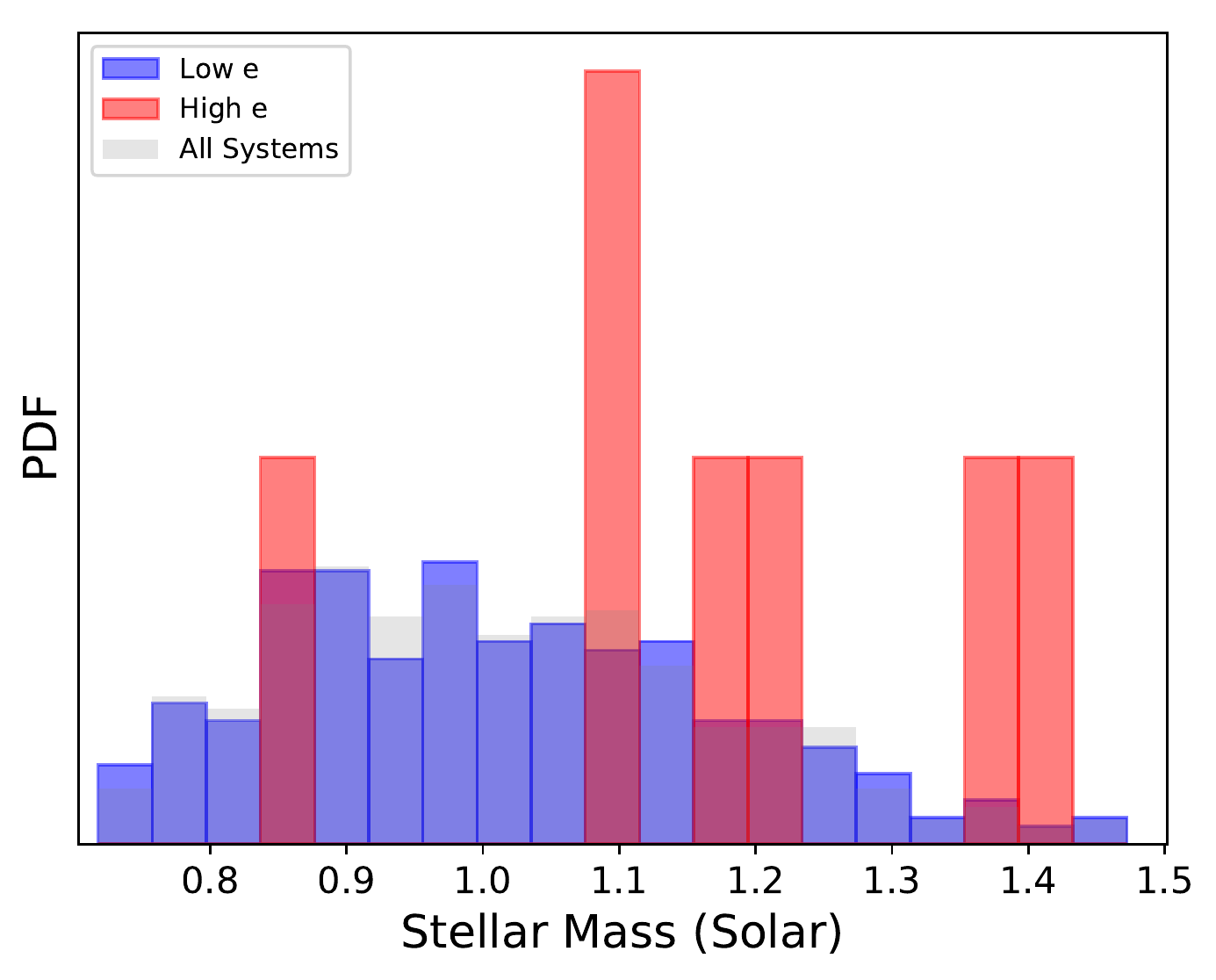}
}
\centerline{
\includegraphics[scale=0.4]{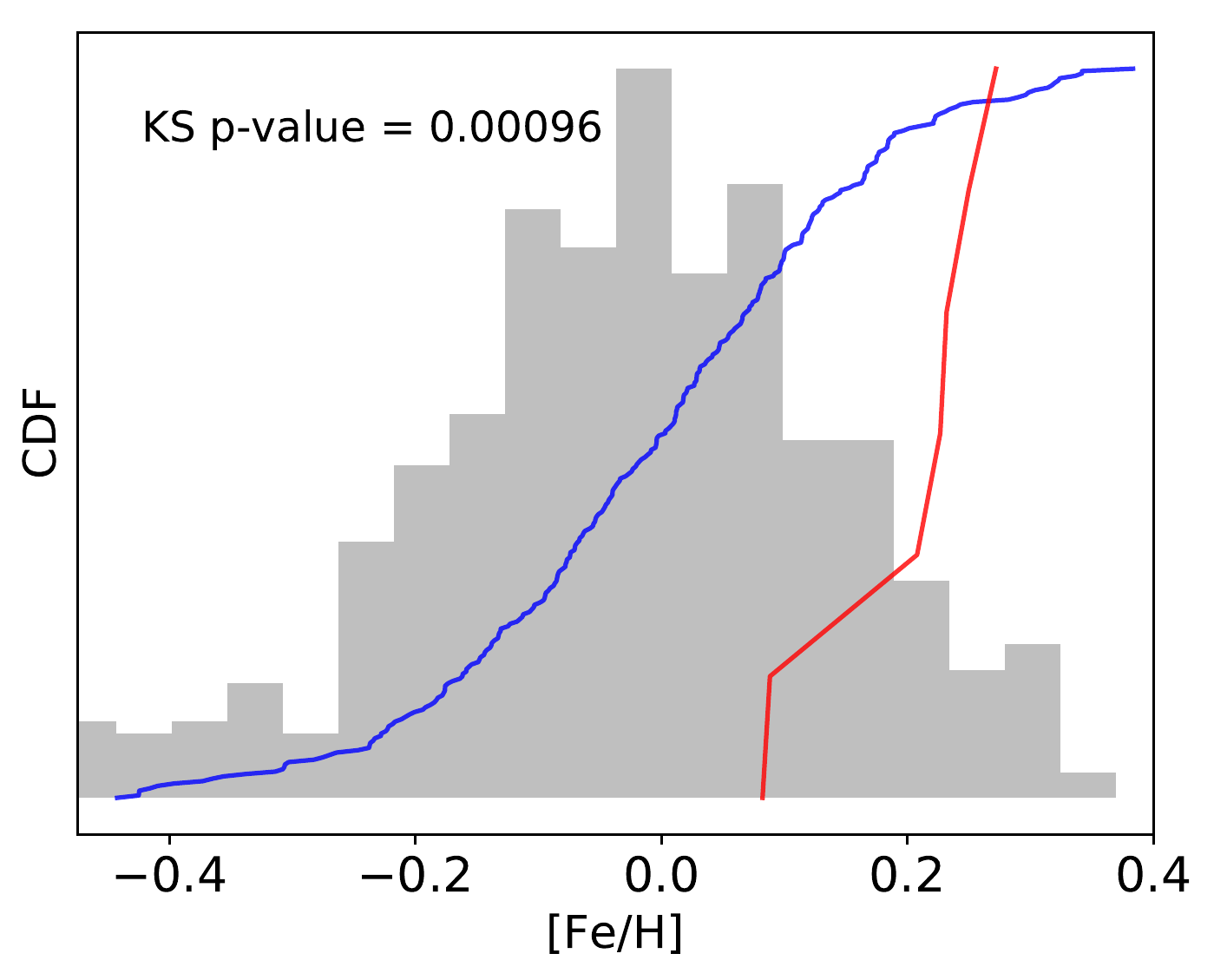} 
\includegraphics[scale=0.4]{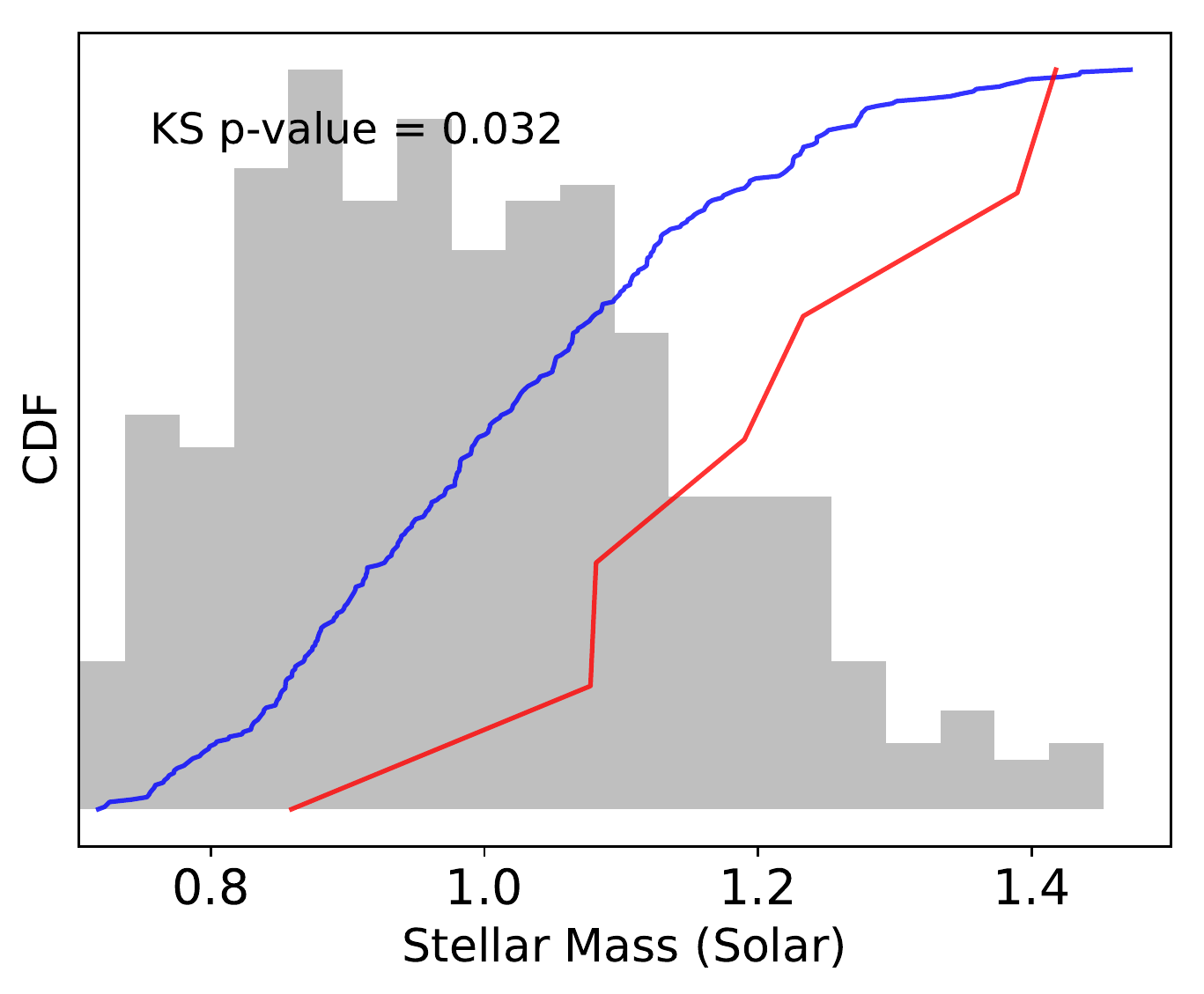}
}
\caption{The cumulative probability distribution functions (CDFs) of the host star [Fe/H] and mass of the population of singly transiting systems which support low eccentricities (blue) and the 5\% of systems which most strongly favor high eccentricities (red). The PDF of the entire sample is shown in gray. 
}
\label{fig:singlesfeh}
\end{figure*}

\subsection{Eccentricity Sub-Population Comparisons}

The presence of a few individual planets out of the 439 considered that determine the lower-bound of the eccentricity distribution via their long durations makes subdividing the population into bins of stellar or planetary properties to identify trends difficult. If the bins are even moderately small, it is probable that some bins contain none of these long-duration planets, resulting in a wildly oscillating $\sigma_e$ distribution between $\sim$0 and $\sim$0.2 as a function of the property. Therefore we adopt an approach which makes use of the previous division into high and low-eccentricity planets and is agnostic to the functional form of the eccentricity dependence. We divide the systems into high and low eccentricity groups for the best-fit $\sigma_{e,low}$ and $\sigma_{e,high}$ described above and then compare the properties of the 5\% of systems which most strongly favor high eccentricity with all of the systems in the low-eccentricity group. This method is adopted for two reasons. First, if we would compare every system in both groups, any differences between the two populations may be strongly diluted by systems which only slightly favor the high eccentricity group due, for instance, to circular systems at the high end of the impact parameter distribution whose short transit durations very modestly ($<1$$\sigma$) favor a high eccentricity distribution. By taking only the top 5\%, we restrict our sample to systems which are very likely to be eccentric, while still having enough systems to make a meaningful comparison. Second, we include the entire low-eccentricity population to get a good sense of the underlying property distribution for comparison. We consider the properties $M_\star$, $R_\star$, $T_\mathrm{eff}$, [Fe/H], $P$, and $R_p$.

This leaves only 7 planets in the high eccentricity group, however each is highly likely to be eccentric. Further, this a sufficient sample for comparison to the $\sim$300 in the low-eccentricity distribution via Kolmogorov-Smirnov two-sample tests, which take into account the number of samples in each distribution. However, we caution against over-interpretation of these results due to the small number of objects involved. For a threshold of statistical significance, we initially select a nominal p-value of 0.01 (2.6-$\sigma$ equivalent detection). 
We also consider the problem of multiple hypothesis testing \citep[see, e.g.,][]{miller1981simultaneous}, which can be alleviated via a Bonferroni Correction \citep{20001561442,miller1981simultaneous}. Dividing the threshold p-value by the number of tests performed, we compute a new threshold of $0.01/6 = 0.002$. Correlations between stellar variables considered \citep[e.g., {[}Fe/H{]}-$M_\star$;][]{2003AA...398..363S,2010PASP..122..905J} would require a less harsh reduction in the p-values \citep[e.g.,][]{sidak1968}, so this correction is conservative. The resulting 0.002 threshold is as conservative as generally recommended for evidence of new effects (even when assuming conservative prior odds of 1:10--0.005, \citealt{43d29956f6b342b6969a4c7bf165d130}; when considering multiple tests and reproducibility--0.003, \citealt{bs1987}; from a Bayesian testing perspective--0.001-0.005, \citealt{2013PNAS..11019313J}; and as commonly used as first evidence for a particle in physics--0.003).
It corresponds to a Bayes Factor of $>$50 compared to the null hypothesis \citep[a significant detection;][]{2013PNAS..11019313J}, and is equivalent to a $>$3-$\sigma$ individual detection. 

We find that the high eccentricity planets do not significantly differ in $P$ or stellar $T_\mathrm{eff}$ from low eccentricity planets via a Kolmogorov-Smirnov two-sample test\footnote{We note that no planet in the top 10\% of the eccentricity distribution has $\log_{10} P  < 0.5$  in units of days, consistent with theories of tidal dissipation \citep{1996ApJ...470.1187R,2010EAS....42..411R}.}. A trend for high eccentricity planets to prefer larger radii than low eccentricity planets is seen at the p = 0.045 level (see Fig.~\ref{fig:singlesall}). We caution against over-interpretation of this result because of the low significance (p-value outside our threshold) and since we are only taking the systems with the most significant eccentricities. It is possible this trend is due to the lower S/N of transits of very small planets, which may have greater uncertainties in transit duration and thus are less likely to strongly suggest a non-zero eccentricity. However, we do not find a significant correlation between planet detection S/N and eccentricity (Fig.~\ref{fig:singlesall}). 

We find that high eccentricity planets are preferentially found around high metallicity ($[Fe/H] > 0$; Fig.~\ref{fig:singlesfeh}). This may in principle be influenced by the marginally larger than expected radii of the high-eccentricity planets as discussed above, because a stellar metallicity-planet radius correlation has been shown to exist \citep{2014Natur.509..593B,2018AJ....155...89P}. However, this sample includes only 1 planet with $R>4 R_\oplus$, and the same trend is observed (KS test p value = 0.0016) when it is removed and the sample restricted to $< 4R_\oplus$ where metallicity is thought to be only weakly dependent on planetary radius \citep{2014Natur.509..593B}. Thus we rule out any radius dependence as the cause of the observed metallicity-[Fe/H] correlation. There is also a slight preference for eccentric planets to have high mass host stars (Fig.~\ref{fig:singlesfeh}), but it is not formally significant. We note that the correlation coefficient, $\rho$, between stellar mass and metallicity in our sample is 0.3. Thus the slight mass preference may be due to the larger number of metal-rich stars at high mass, but high mass stars could not be the cause of the observed eccentricity-[Fe/H] correlation. Additionally, the preference for higher $M_\star$ decreases in significance when considering only planets $< 4R_\oplus$ (p value = 0.064; outside our threshold). Further work is needed to validate both trends, and the mass trend in particular is not confidently observed.

\subsection{Additional Validation}

\begin{figure}
\centerline{
\includegraphics[scale=0.6]{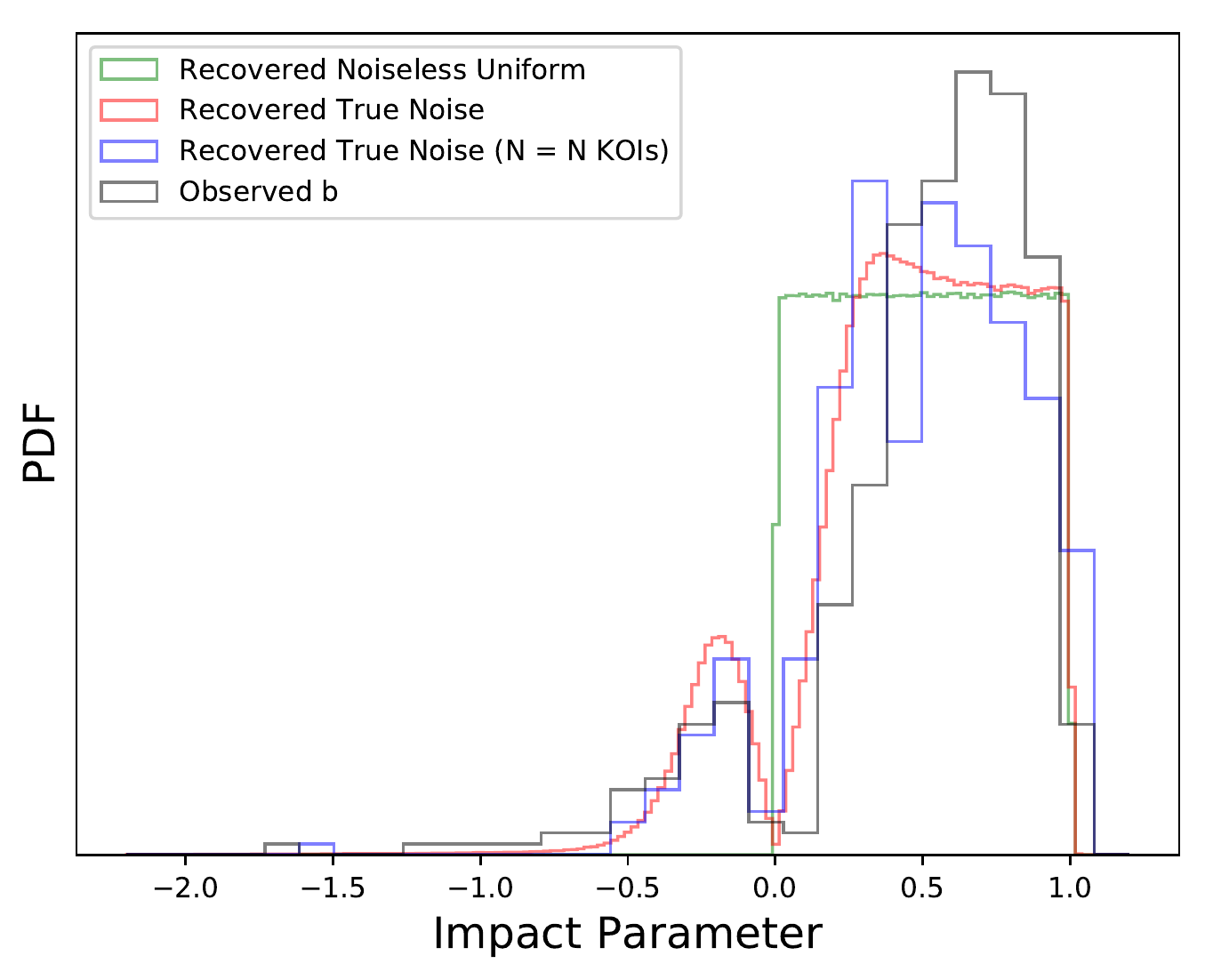}
}
\centerline{
\includegraphics[scale=0.6]{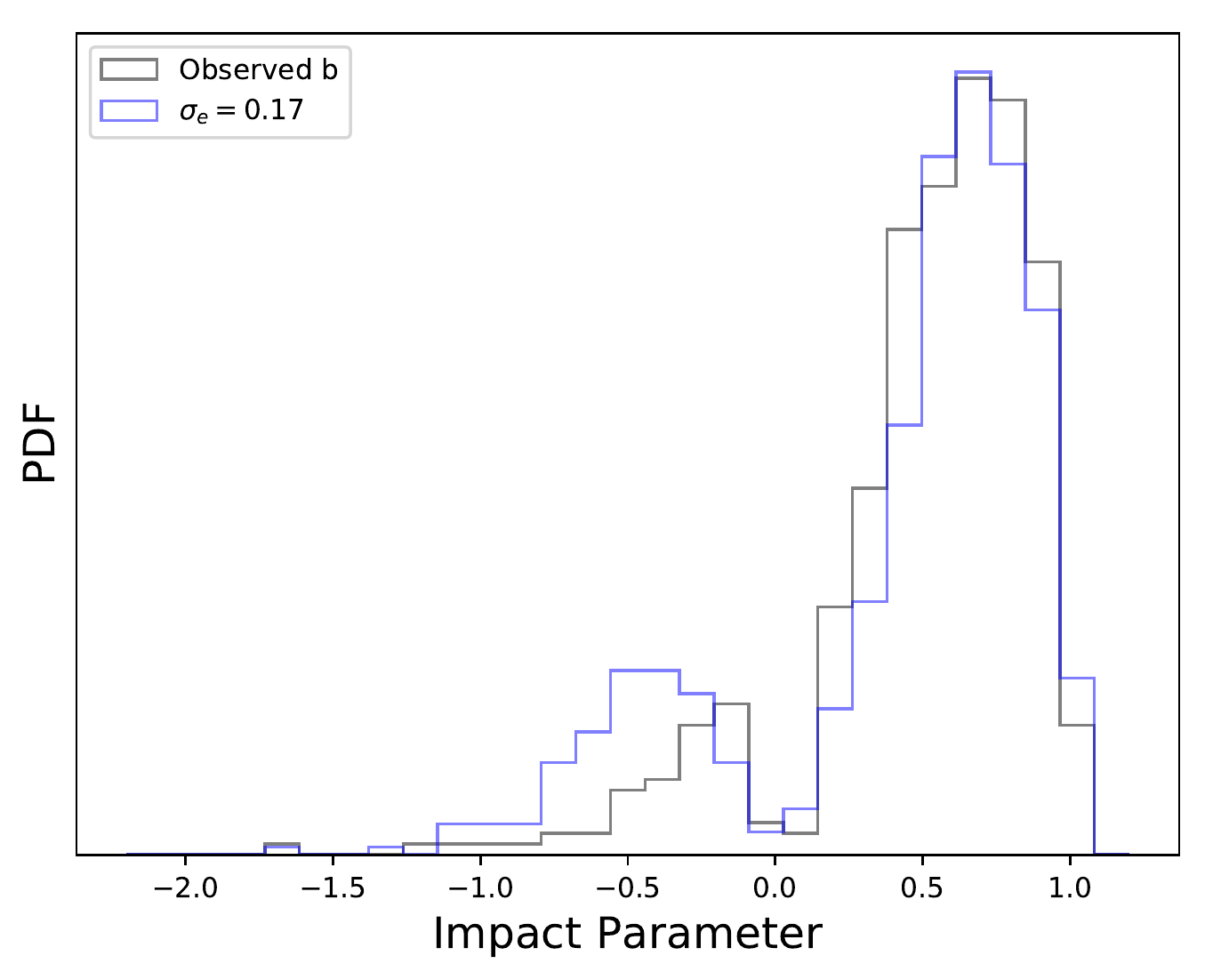}
}
\caption{
\emph{Top:} Injected and recovered uniform $e=0$ impact parameter distributions compared to the observed $b$ distribution for singly transiting systems. We show the results of recovering an injected uniform $b$ distribution with no noise (green), which matches the injected distribution almost perfectly. We also show two different $e=0$ injection and recovery samples: one with many iterations to ensure the distribution is smooth and well sampled (red), and a second with as many samples as the observed single-planet KOIs so that the level of variation due to Poisson noise can easily be seen (blue). The agreement between the blue $e=0$ sample and the black observations is visibly poor. \emph{Bottom:} A realization of the distribution of impact parameters if the population has $\sigma_e = 0.17$ (blue), complete with S/N cutoffs as described in the methods compared to the observed distribution of $b$s in black. These distributions match more closely than the $e=0$ example, but are still not identical, suggesting a two-population model.  
}
\label{fig:bdistsingles}
\end{figure}

For an intuitive understanding of the comparison between the observed durations and the eccentricity, we show the distribution of impact parameters implied from the measured \Kepler durations assuming a population with $e=0$ in Fig.~\ref{fig:bdistsingles}. We then compare it with simulations of the distributions recovered by drawing from the stellar and orbital property uncertainties, and injecting a uniform distribution of impact parameters. This allows us to see how the addition of the uncertainties changes the nearly uniform distribution of impact parameters to a double-peaked distribution disfavoring $b\approx0$. In order to take into account longer than expected transit durations given the $R_\star$ value and a circular planet's velocity, we analytically extend the formula for the $e=0$ impact parameter ($b_a$) as 
\begin{equation} \begin{split}
 b_1 & = \bigg(1+ \frac{R_p}{R_\star}\bigg) - \bigg(\frac{a}{R_\star} \sin\big(\frac{\pi d}{P}\big)\bigg) \\
 b_a & = \sgn(b_1) \sqrt{|b_1|}.
\end{split} \end{equation}
Thus transit durations that are longer than theoretically possible for  an ($R_\star$, $v_p$) pair due to either eccentricity or measurement error are treated as negative impact parameters and the distribution is thus real for all possible input values. We note that although $b$ is isomorphic to duration, the transformation is nonlinear and therefore the introduction of noise in the duration (and other parameter) measurements will not simply broaden an initial uniform distribution of impact parameters but also distort it. For $b$ in [0,1+$R_p/R_\star$), $|\frac{\partial b}{\partial d}|$ is strictly increasing and thus causes a bias towards high $b$. In other words, the density of durations as a function of $b$ grows as $b$ increases away from 0 and thus random noise preferentially biases $b$s to higher values. This effect increases with the level of noise. Additionally, noise may cause negative best-fit $b$ values (which would create a pile-up at $b=0$ if negative $b$ was not allowed). We emphasize this here as it has caused some confusion in the past \citep[c.f.,][]{0004-637X-778-1-53}. Fig.~\ref{fig:bdistsingles} shows that the single planet systems have both outliers at large negative values and a surplus of apparent high impact parameters -- both indicators of eccentricity due to transit near apocenter and pericenter respectively \citep{2008ApJ...679.1566B}. The routine described in \S~\ref{sec:methodssingles} can be restated as a method of measuring how well the observed and theoretical $b$ distributions match as a function of population eccentricity, while taking into account the uncertainties and observational bias which causes a decline in detection at very high $b$. The multiple planet case is more complicated due to the non-uniform underlying $b$ distribution from mutual inclination constraints, so we do not show a similar simplified figure for such systems. 

\begin{figure*}
\centerline{
\includegraphics[scale=0.5]{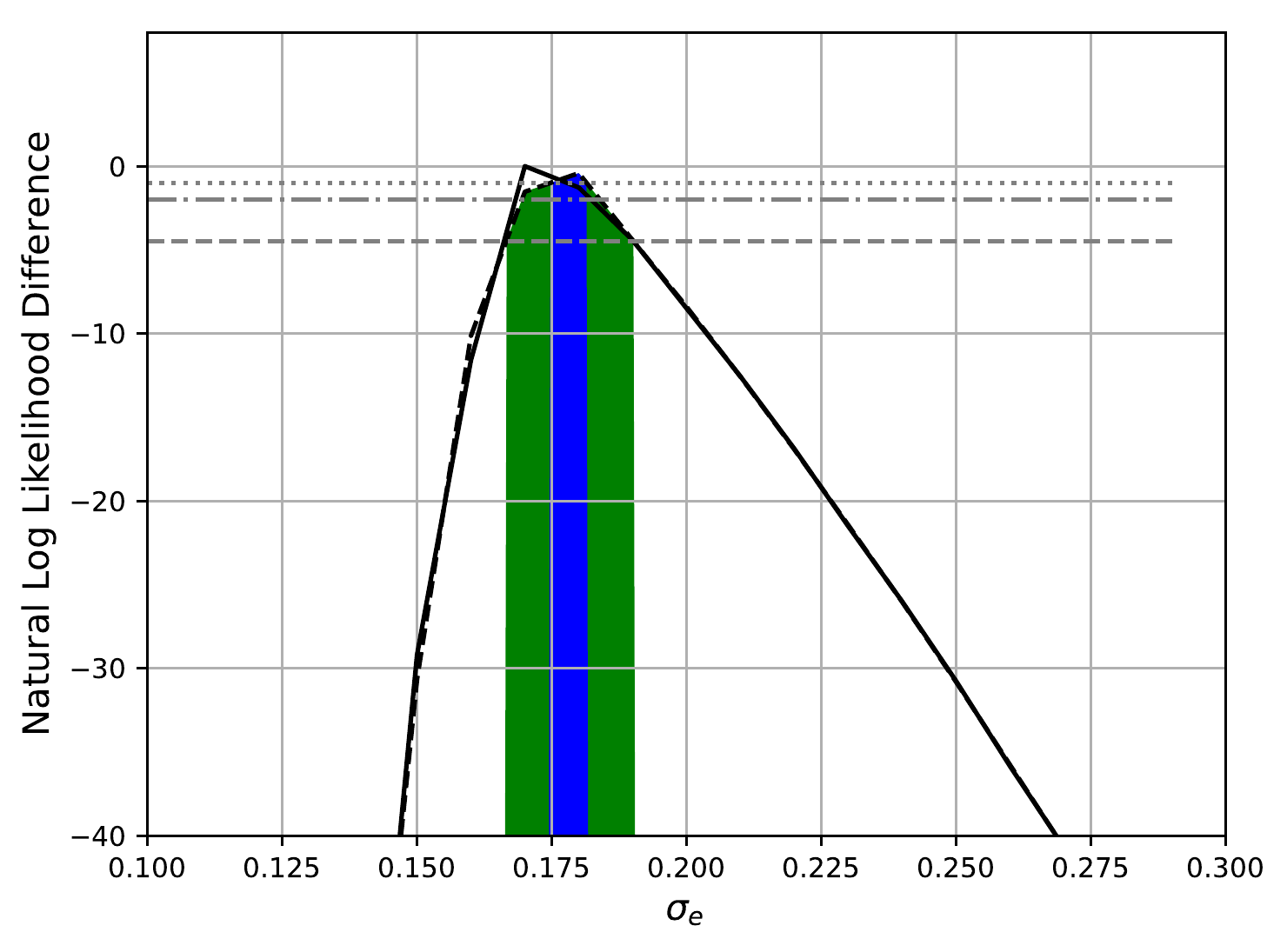}
\includegraphics[scale=0.5]{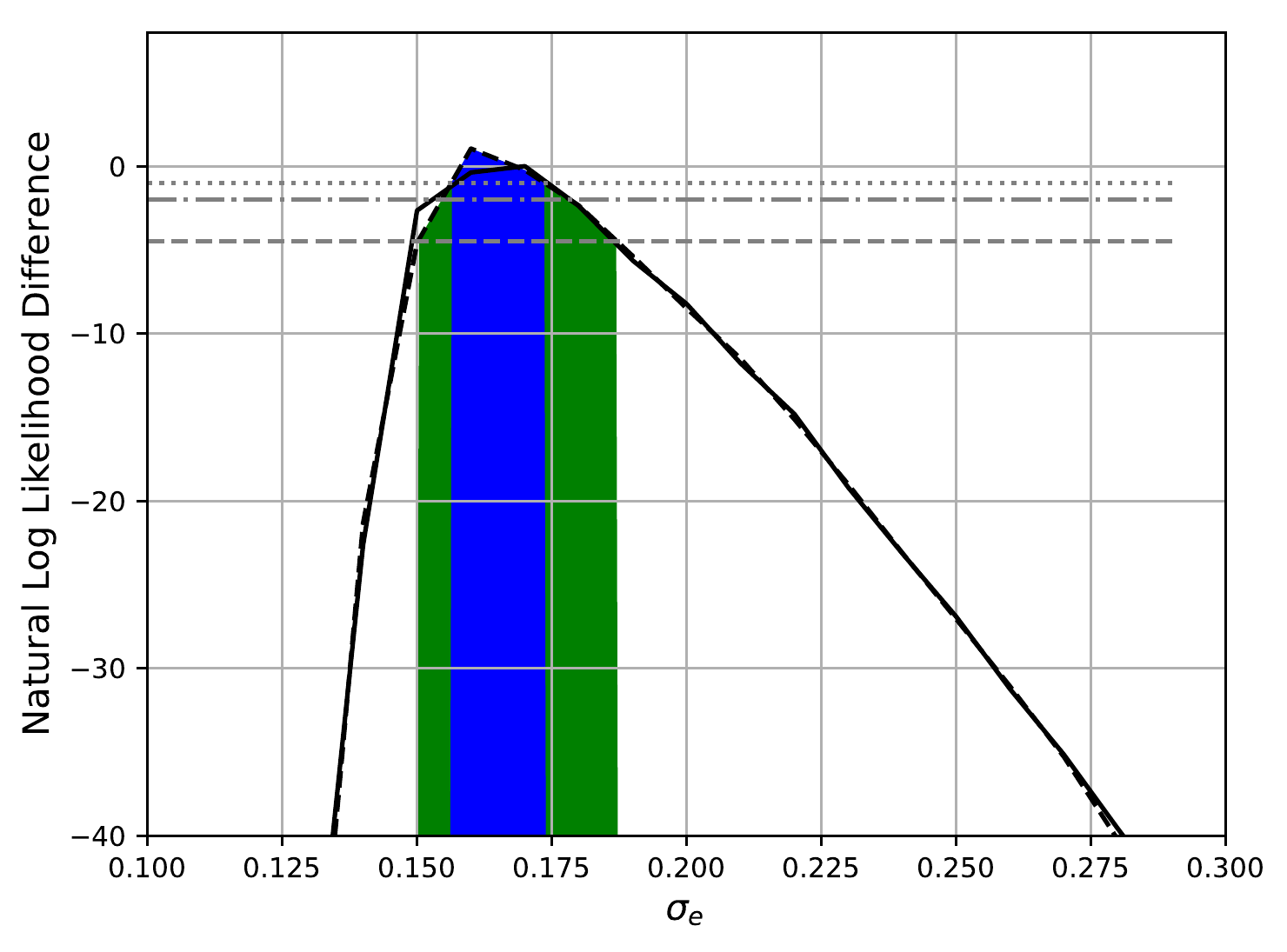}
}
\caption{ Singly transiting planet eccentricity distribution fits with modified stellar parameters. Similar to Fig~\ref{fig:allfit}, we show the likelihood as function of $\sigma_e$ for a test where all stellar radii are shrunk by 3\% (left) and increased by 3\% (right). This is insufficient to drastically change the conclusion of the high $\sigma_e$ for the single transiting planet systems. 
}
\label{fig:starvar}
\end{figure*}

The median radius uncertainties on the CKS-\Gaia stars are $\approx3\%$, and have been shown to be accurate compared to asteroseismology samples \citep{2018AJ....156..264F}. Nevertheless we conduct a test to determine how much unaccounted for systematic biases in the stellar properties would change our results. We re-fit the eccentricity distributions of the singles using a new set of stellar properties, where the stellar radii are all changed by 3\% higher and lower with the reported uncertainties unchanged. The results do change the greatest likelihood $\sigma_e$ and uncertainties, but still strongly rule out a $\sigma_e$ near zero (see Fig.~\ref{fig:starvar}). Since the transit duration is more weakly dependent on stellar mass, systematic biases are even less important there. Any biases for other stellar parameters would not affect the observed trends since the relative distributions of, for instance metallicity, between the two populations are measured rather than any absolute values being used. Thus we confirm that even if the stellar properties are systematically wrong by several $\sigma$ in either direction, the qualitative results of our study will stand. 

We also consider systematics in the \Kepler data. We compare transit durations reported in Data Release 24 \citep[DR24;][]{2016ApJS..224...12C} and Data Release 25 \citep[DR25;][]{2016AJ....152..158T}. The majority agree to within 1-$\sigma$ and an approximately normal distribution is generated by examining $(T_{dur,DR25}-T_{dur,DR24}))/\sigma_{dur,DR25}$ with $\sigma = 1.05$, excluding a few outliers. 

\section{Multiple Planet Results}

\begin{figure}
\centerline{
\includegraphics[scale=0.6]{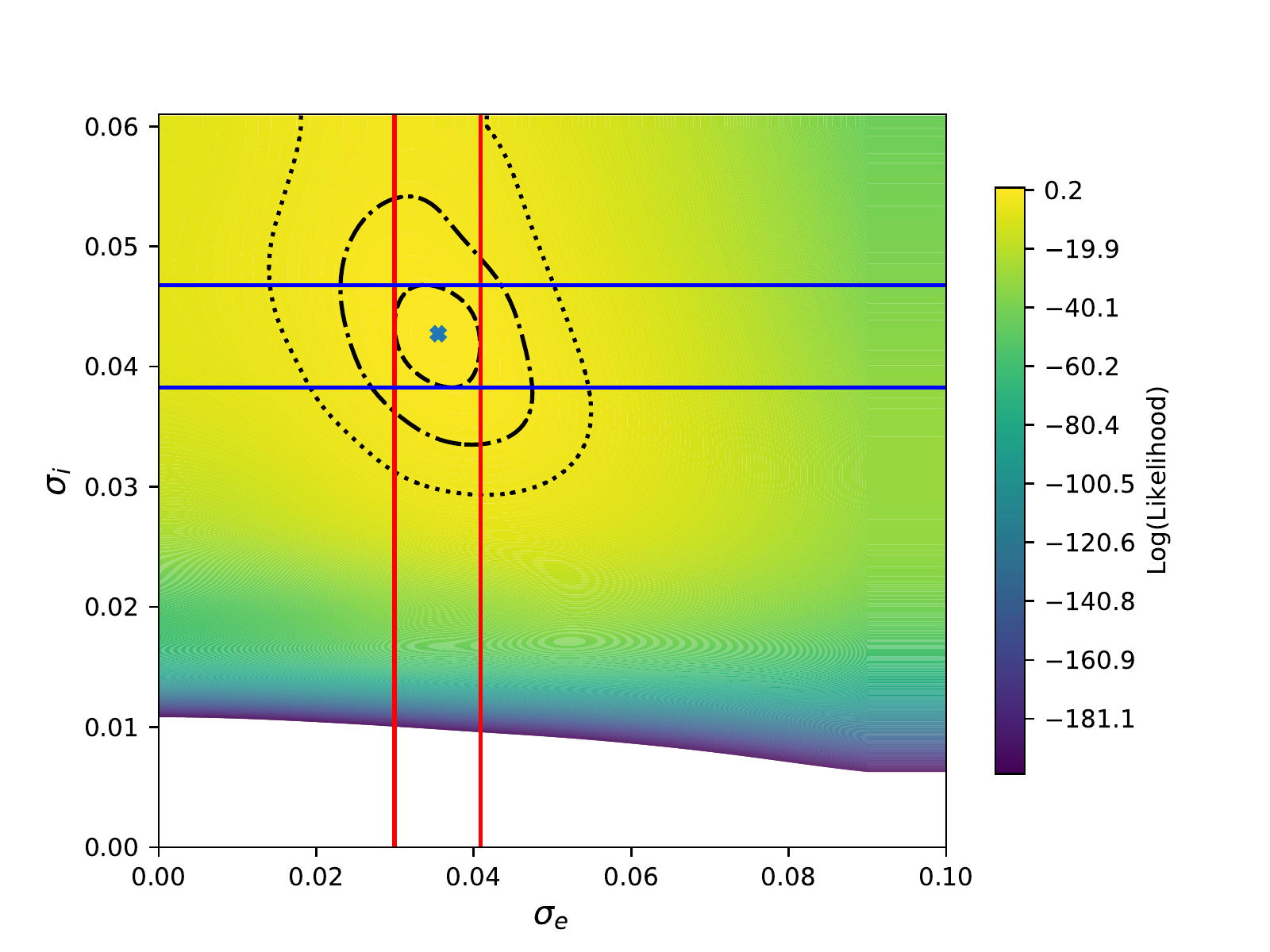}
}
\caption{The contours of natural log likelihood for fits of $\sigma_e$ and $\sigma_i$ values for multiply transiting \Kepler systems are shown with dashed, dot-dashed, and dotted lines indicating 1-, 2-, and 3-$\sigma$ confidence levels. The red and blue lines show the 1-$\sigma$ confidence interval projected along the axes for clarity. This figure is similar to Fig.~\ref{fig:allfit}, except we must also consider the mutual inclination among planets in addition to just the eccentricity distribution in order to fit the transit durations. A comparison to the single-planet $\sigma_e$ distribution is shown in Fig.~\ref{fig:allfit} by taking a horizontal slice of the likelihood distribution at $\sigma_i = 0.043$. 
}
\label{fig:multis2d}
\end{figure}

The results of the transit duration simulations accounting for mutual inclination among multiply transiting planetary systems outlined in \S \ref{sec:methodsmultis} indicate that the population of planets in multiply transiting systems have low eccentricities ($\sigma_e = 0.0355\pm0.012$ at the 2-$\sigma$ level). We show a contour plot of the data with a cubic spline smoothing in Fig.~\ref{fig:multis2d}. The mutual inclination distribution is found to have $\sigma_i = 2.45^{+0.65}_{-0.53}$ degrees at the 2-$\sigma$ level.  
Similar to the single-planet case, we consider individual systems whose transit durations rule out low eccentricities. Since the final multiplanet sample consists of 870 KOIs, we look for systems which prefer non-zero $\sigma_e$ at greater than 3.7$\sigma$, as this leads to the expectation of fewer than 1 false positive. 
No systems reach this cutoff, suggesting a uniformly low eccentricity distribution. 

We may still divide the KOIs into groups that prefer high and low eccentricity values as we did with the single transiting planets, even if systems do not individually strongly prefer one or the other. In addition to the parameters considered for the singles, we also consider the period ratio of the nearest neighboring planet ($P_{rat}$), the size of the largest planet in the system in Earth radii ($R_{max}$), and the number of planets in the system ($N_{pl}$). However, we find no statistically significant differences between the distributions of any of these parameters using the same methodology as for the singles at the p = 0.05 level.  This is perhaps unsurprising as it is far more difficult to separate high and low eccentricity planets in this sample.

\section{Summary and Discussion}

The most important insight this eccentricity population study provides is that high $e$ planets of $1.4-4R_\oplus$ size are preferentially found around metal-rich stars. Taken together with previous work on larger planets \citep{2013ApJ...767L..24D,2018ApJ...856...37B}, these results suggest that eccentric planets of all sizes are preferentially found around super-Solar metallicity stars. We also confirm previous results that the single transiting planet systems in \Kepler are drawn from a broader range of eccentricities than the multiply transiting systems. The Rayleigh $\sigma_e$ values we find are slightly lower than reported for the singles ($\bar{e}\approx0.3$) in X16, but agree with being inconsistent with $e\approx0$. We note that we also performed a validation study using only the CKS spectral data, which resulted in a slightly higher $e$ distribution for both the single and multiply transiting systems. This suggests that the improved accuracy provided by the \Gaia calibrations on the stellar properties \citep{2018AJ....156..264F}, may explain the slight discrepancy between the results. On the other hand, our multiplanet $\sigma_e$ agrees well with X16's results as well as those found by \citet{2015ApJ...808..126V}, suggesting low eccentricities around the majority of multiply transiting \Kepler systems.

A well-known correlation exists between the existence of giant planets and a star's metallicity \citep{2005ApJ...622.1102F} and mass \citep{2010PASP..122..905J}. It is notable that these are the two characteristics which are most strongly correlated with eccentricity in the single-transiting planet case. High stellar mass is associated with more massive disks \citep{2013ApJ...771..129A}, which may encourage giant planet formation \citep{2008ApJ...673..502K}, but also decreased lifetimes which may inhibit it \citep{2015AA...576A..52R}. However, high metallicity environments promote both more robust planet formation and longer disc lifetimes increasing both the frequency of giant planets \citep{2010ApJ...723L.113Y,2010MNRAS.402.2735E} and perhaps Earth to sub-Saturn planets (\citealt{2015AJ....149...14W,2018AJ....155...89P}, but c.f. \citealt{2015ApJ...808..187B}). These giant planets may interact with compact inner planet systems to increase eccentricity while decreasing multiplicity or apparent multiplicity due to greater mutual inclinations \citep{2017AJ....153..210H,2017MNRAS.467.1531H,2018MNRAS.478..197P}, whereas typical multiplanet systems may not reach high eccentricities and mutual inclinations via self-excitation \citep{2016MNRAS.455.2980B}. If this hypothesis is correct, a search for giant planets around stars hosting an apparent single eccentric planet should recover companions at a high rate. 

More work and greater statistical certainty is required to disentangle the different influences on formation with the distributions of planetary eccentricities. We therefore look forward to future transit surveys such as TESS \citep{2014SPIE.9143E..20R} and PLATO \citep{2014ExA....38..249R} and their follow-up campaigns, which will provide many more transiting planet durations to validate the planetary eccentricity -- stellar metallicity relationship.

\acknowledgements

We thank the Kepler and Gaia teams for years of work making these precious datasets possible.  
We are grateful to the time assignment committees of the University of Hawaii, the University of California, the California Institute of Technology, and NASA for their generous allocations of Keck observing time that enabled this large project.  
This research has made use of NASA's Astrophysics Data System, the Exoplanet Orbit Database, and the Exoplanet Data Explorer at exoplanets.org.  
A.W.H. acknowledges NASA grant NNX12AJ23G. L.M.W. acknowledges support from the Beatrice Watson Parrent Fellowship.
The authors wish to recognize and acknowledge the very significant cultural role and reverence that the summit of Maunakea has long had within the indigenous Hawaiian community. We are most fortunate to have the opportunity to conduct observations from this mountain.

\end{document}